\documentclass[12pt, letter]{article}

\usepackage{amsmath}
\usepackage{amssymb}
\usepackage{fullpage}
\usepackage{color}
\usepackage{xspace}
\usepackage{natbib}
\usepackage{graphicx}

\input{macro_wjournal_wxspace.inp}

\begin{document}

\begin{center}
\noindent{\large Astro 2020 APC White Paper}\\
{\Large\bf A Space-based All-sky MeV $\gamma$-ray Survey\\ \vspace{2mm} with the Electron Tracking Compton Camera}
\end{center}

\noindent{\bf Thematic Activity/Project/state of the Profession Consideration Area:}\\
A sensitive survey of the MeV $\gamma$-ray sky is needed to understand important astrophysical problems
such as $\gamma$-ray bursts in the early universe, progenitors of Type Ia supernovae, and the nature of dark matter.
However, the study has not progressed remarkably since the limited survey by COMPTEL onboard CGRO in the 1990s.
Tanimori et al.\ have developed a Compton camera that tracks the trajectory of each recoil electron 
in addition to the information obtained by the conventional Compton cameras, leading to superior imaging.
This Electron Tracking Compton Camera (ETCC) facilitates accurate reconstruction of the incoming direction of each MeV photon from a wide sky
at $\sim$degree angular resolution and with minimized particle background using trajectory information.
The latest ETCC model, SMILE-2+, made successful astronomical observations during a day balloon flight in 2018 April
and detected diffuse continuum and 511 keV annihilation line emission from the Galactic Center region at a high significance
in $\sim$2.5~hours.
We believe that MeV observations from space with upgraded ETCCs will dramatically improve our knowledge of the MeV universe.
We advocate for a space-based all-sky survey mission with multiple ETCCs onboard and detail its expected  benefits.\\

\noindent{\bf Principal Authors:}\\
Kenji Hamaguchi$^{1,2, *}$, CRESST II NASA/GSFC \& UMBC, kenji.hamaguchi@umbc.edu\\
Toru Tanimori$^{3}$, Kyoto University, tanimori@cr.scphys.kyoto-u.ac.jp\\
Atsushi Takada$^{3}$, Kyoto University, takada@cr.scphys.kyoto-u.ac.jp\\

\noindent{\bf Co-authors:}\\
John F. Beacom, Ohio State University, beacom.7@osu.edu\\
Shuichi Gunji, Yamagata University, gunji@sci.kj.yamagata-u.ac.jp\\
Masaki Mori, Ritsumeikan University, morim@fc.ritsumei.ac.jp\\
Takeshi Nakamori, Yamagata University, nakamori@sci.kj.yamagata-u.ac.jp\\
Chris R. Shrader, CRESST II NASA/GSFC \& Catholic University, chris.r.shrader@nasa.gov\\
David M. Smith, University of Calfornia Santa Cruz, dsmith8@ucsc.edu\\
Toru Tamagawa, RIKEN, tamagawa@riken.jp\\
Bruce T. Tsurutani, NASA/JPL, bruce.t.tsurutani@jpl.nasa.gov\\

\noindent{\bf Contact Addresses of Principal Authors:}\\
1: Code 662.0, NASA/GSFC, Greenbelt, MD 20771, USA\\
2: Department of Physics, University of Maryland, Baltimore County, 1000 Hilltop Circle, Baltimore, MD 21250, USA\\
3: Department of Physics, Kyoto University, Kitashirakawa-Oiwakecho, Sakyo-ku, Kyoto 606-8502, Japan

\clearpage
\pagenumbering{arabic}
\setcounter{page}{1}

\section{MeV $\gamma$-ray Sky Needs a Revolutionary Instrument}

Observational astronomy has made incredible advances over the past few decades throughout nearly 
the entire electromagnetic spectrum, with the important exception of the 0.1$-$100 MeV band.
This energy band is a window of crucial high energy particle interactions, 
such as electron-positron annihilation, radioisotope decay lines (e.g., $^{26}$Al, $^{60}$Fe), pion decay emission
as well as inverse Compton emission from relativistic particles.
This band is suspected to hold key information on current astrophysical conundrums, such as 
$\gamma$-ray bursts in the early universe, progenitors of Type Ia supernovae \citep[e.g.,][]{Horiuchi2010a}, 
the nature of dark matter, and nucleosynthesis in our Galaxy (see the related science white papers).
The MeV band is also important for understanding high energy activities around Sun, Earth\citep[e.g.,][]{Tsurutani2018a},
and other planets, such as Jupiter.

The first all-sky survey in MeV $\gamma$-rays was performed with the Imaging Compton Telescope (COMPTEL)
onboard Compton Gamma-Ray Observatory between 1991$-$2000 \citep{Schoenfelder1993a}.
The telescope covered 1 steradian of the sky between 0.8$-$30~MeV, to a sensitivity 
of 0.1~Crab near 1 MeV over $\sim$8 years, but detected only a few dozen persistent sources.
The IBIS and SPI instruments onboard \INTEGRAL have performed MeV all-sky surveys since 2002
with better sensitivities than COMPTEL below 1 MeV.
IBIS has detected 132 sources above 100~keV in 11 years \citep{Krivonos2015a}.
SPI did not detect many sources, 
but SPI's excellent spectral resolution improved measurements of 511~keV annihilation lines \citep{Beacom2006a}
and $^{26}$Al and $^{60}$Fe nuclear lines from our galaxy, and detected nuclear $\gamma$-rays from a type-Ia supernova 
(2014J) for the first time \citep{Churazov2014a,Diehl2014a}

However, these results suffer huge cosmic ray induced instrumental background
and therefore their detections are limited to very bright sources.
This is because these instruments cannot localize individual photons nor 
distinguish photon events from particle background events.
MeV $\gamma$-ray photons mostly interact with materials via Compton scattering.
No optics have been successfully developed to focus MeV $\gamma$-ray photons.
Both IBIS and SPI instruments use a coded mask technology,
which decodes shadow mask images cast by bright sources,
but this method needs to collect many photons to beat statistics, calibration uncertainty and background
and so only works for very bright sources.

\begin{figure}[t]
\includegraphics[width=6.5in]{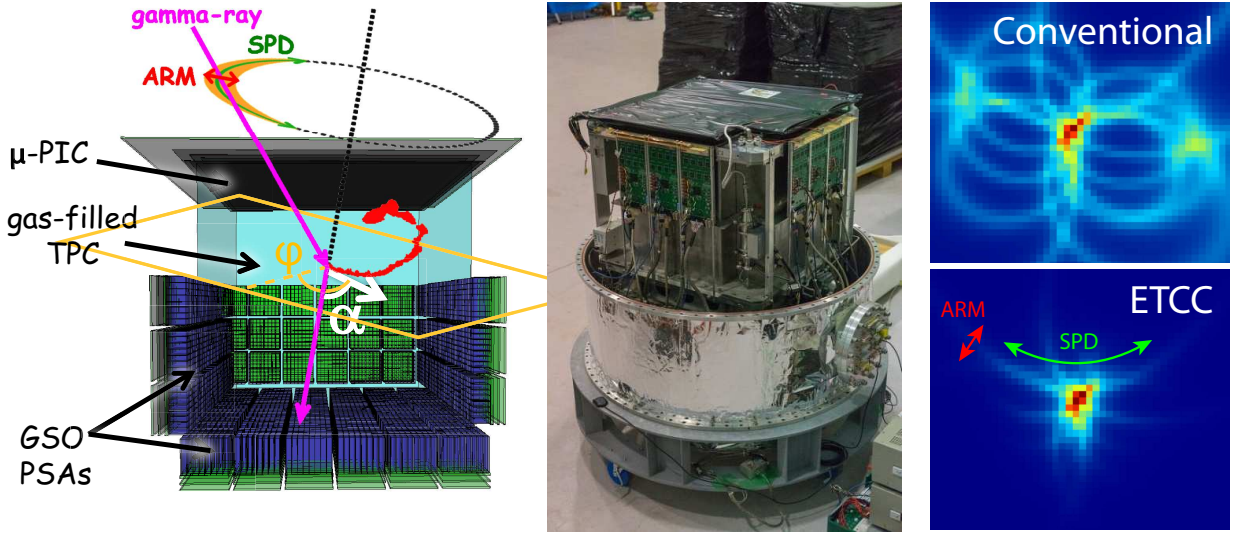}
\caption{
{\it Left} --- Schematic view of SMILE-2+ 30 cm-cubic ETCC \citep{Tanimori2015a}.
A micro-pattern gas detector ($\mu$-PIC), which consists of 400 $\mu$m pitch pixels, is installed at the TPC top, 
of which anodes and cathodes are connected via strips to provide two-dimensional charge tracks.
{\it Middle} --- Photograph of SMILE-2+ flight model instrument.
{\it Right} --- Point source images with the conventional Compton camera ({\it top}) and with SMILE-2+ 
ETCC ({\it bottom})
\citep{Mizumura2014a}.
\label{fig:ETCC}
}
\end{figure}

COMPTEL used the Compton scattering process to constrain the direction of each incoming photon.
This type of instrument, a conventional Compton camera, consists of two main modules,
the scattering chamber and calorimeter arrays.
An incoming $\gamma$-ray interacts with an electron in the scattering chamber 
and scatters via the Compton process.
The recoiled electron deposits energy around the scattering location,
while the scattering $\gamma$-ray photon hits a calorimeter.
Their position and energy information are combined to solve the scattering angle 
and the energy of the incoming photon via the Compton scattering equation.
The solution constrains the incoming direction of each photon within an annular region called ``event circle"
around the scattering photon direction (see Figure~\ref{fig:ETCC} {\it left}).
This method was also used for the recent COSI balloon experiment \citep{Chiu2017a}.

Conventional Compton cameras produce many annuli in an image, which provide probable incoming directions of individual photons
(Figure~\ref{fig:ETCC} {\it right}).
Intersected positions are likely $\gamma$-ray sources that emit MeV photons.
Sources are detected with image deconvolution techniques such as ``Maximum Entropy Method",
which, however, cannot recover faint sources nor extended sources obscured by bright sources.
These methods also produce artifacts even for bright sources.
This problem is most severe in observations from space under strong particle radiation
as background events smear these peaks.
Most classical Compton cameras
rely on veto counters for rejecting particle events, but these counters themselves produce 
background emission and/or particles, and
therefore, background cannot be efficiently removed.

In Compton scattering, the momentum of an incoming photon on the plane normal to the scattering direction 
is fully given to the recoil electron.
By measuring each electron recoil track, the incoming photon direction is further constrained to a small area, 
i.e. the Compton scattering kinematics can be completely solved as the conical degeneracy inherent in the event reconstruction process is removed.
Tanimori et al.\ at Kyoto University in Japan have developed a Compton camera that tracks recoil electrons of each Compton scattering
with a gaseous micro Time Projection Chamber (TPC) \citep{Takada2011a,Tanimori2017a}
This Electron Tracking Compton Camera (ETCC) reconstructs the incoming direction of each MeV $\gamma$-ray photon within degrees
and also substantially excludes particle background events with additional ``redundant" pieces of information of electron recoil tracks.
The current model, SMILE-2+, had a successful balloon flight in April 2018 in Australia \citep{Takada2016a,Nakamura2018a}
and detected diffuse MeV $\gamma$-rays (0.2$-$2~MeV) with the 511 keV electron-positron pair production line
around the Galactic Center at a significance of $\sim$10$\sigma$ in 2.5~hours of exposure.
ETCC will improve the performance further with a minor upgrade.
In this white paper, we propose a space-based mission with the advanced ETCC to
revolutionize our knowledge of the MeV $\gamma$-ray sky.

\section{Electron Tracking Compton Camera (ETCC)}

\begin{figure}[t]
\includegraphics[width=6.6in]{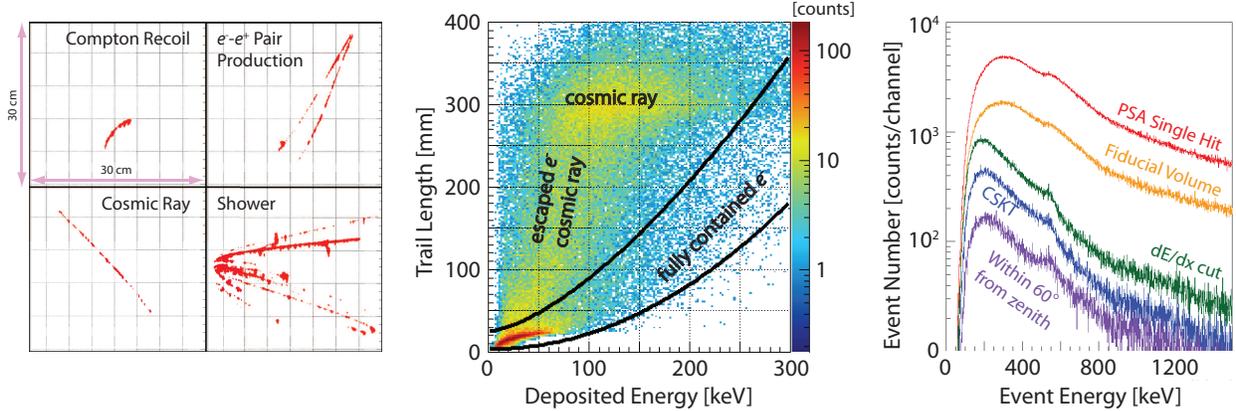}
\caption{
{\it Left} --- Tracks of charged particles originating from different physical processes measured with $\mu$-PIC.
{\it Middle} --- Relation of the deposited energy ($dE$) and the trail length ($dx$) of charged particle events.
Compton Recoil electrons that deposit their whole energy inside the TPC are located within solid black lines (fully contained $e^{-}$).
{\it Right}: Event reduction using multiple screening criteria for data obtained during the SMILE-2+ balloon flight in 2018 April.
CSKT: Compton scattering kinematic test. 
\label{fig:Readout}
}
\end{figure}

\subsection{Design and Performance of the Latest ETCC, SMILE-2+}
\label{subsec:design_performance}

ETCC mainly consists of time proportional chamber (TPC) with a micro pixel gas chamber ($\mu$-PIC),
and pixel scintillator arrays (PSAs), which consist of sets of a GSO scintillator and a photomultiplier tube (PMT),
on 5 sides of the TPC (Figure~\ref{fig:ETCC} {\it left}).
Both modules are placed inside a gas chamber filled with Ar gas at 2 atm.
A $\gamma$-ray photon entering into the TPC interacts with an electron in the Ar gas via the Compton scattering,
and its scattering photon is detected with one of the PSAs.
The electron recoils with a part of the momentum of the incoming photon and ionizes Ar atoms,
producing a track of ionized electrons, which are collected at the TPC top
and read out with $\mu$-PIC with 400 $\mu$m pixel resolution.
The onboard computer matches the PSA event with the TPC electron track.

The total number of ionized electrons provides the initial energy of the recoil electron,
while the starting point of the electron track provides the Compton scattering location.
These results combined with the PSA measurement of the scattering $\gamma$-ray photon
yields the incoming photon energy and the scattering angle through the Compton scattering equation.
The energy measurement errors leave an uncertainty range in the scattering angle, which is called 
the angular resolution measure, or conventionally ARM.
This information constrains the incoming $\gamma$-ray direction to an event circle with an ARM width,
centered at the scattering photon direction (Figure~\ref{fig:ETCC} {\it left}).
These results are what the conventional Compton cameras also provide.

One advantage of the ETCC is that it further refines the incoming ${\gamma}$-ray photon direction.
The initial direction of each electron track holds the vector information of the momentum 
given to the recoil electron from the $\gamma$-ray photon (see Figure~\ref{fig:ETCC} {\it left}).
The incoming $\gamma$-ray direction can be traced back by a projection of this vector onto the plane 
with the event circle, with an uncertainty of this projected angle, called the scatter plane deviation or SPD.
This localizes each incoming photon direction to a small area
and accumulation of $\gamma$-ray photons from a source produces a sharp point-source image 
(Figure~\ref{fig:ETCC} {\it right bottom}).
The ETCC thus provides a well-defined point-spread-function (PSF) and, for the first time, 
MeV $\gamma$-ray imaging that is comparable to optical or X-ray images.
The imaging capability of the ETCC was demonstrated by an absolute intensity mapping of the 
Fukushima nuclear accident site for ground nuclear contamination, as well as
the scattering of MeV $\gamma$-rays from the atmosphere
\citep{Tomono2017a}.
ETCC is selected as the only $\gamma$-ray imaging device for decommissioning Fukushima nuclear reactors
by Nuclear Safety Research Association, a government-funded agency in Japan.

ETCC collects two more ``redundant" physical parameters, which are not necessary for localizing incoming photon directions
but powerful for discriminating particle events.
The first is the trail length.
Compton recoil events show a clear relation between the trail length ($dx$) and the deposited energy ($dE$),
and which is clearly different from those of other (e.g., cosmic ray) events (Figure~\ref{fig:Readout} {\it middle}).
The second is the projection angle of the initial recoil vector onto the event circle plane,
the angle $\alpha$ in Figure~\ref{fig:ETCC} {\it left}.
This angle tests if the electron motion can be due to Compton scattering
(i.e., Compton scattering kinematic test).
By screening events with these parameters,
ETCC can reject most particle background events (Figure~\ref{fig:Readout} {\it right}).

The current SMILE-2+ performance has 
a PSF of 15\DEGREE (FWHM), effective area of 1.5 (0.7)~cm$^{2}$ at 0.511 (1)~MeV
and the sensitive band between 0.2$-$2~MeV (Table~\ref{tbl:roadmap}, Figure~\ref{fig:ETCC_satellite_performance}).\\

The sensitivity declines for $\gtrsim$2~MeV photons because their recoil electrons are
so energetic that they escape from the TPC before losing all of their energy.
These electrons tend to hit a PSA and deposit their remaining energy to it.
By counting these additional PSA (double) hit events,
we should be able to recover high energy photon events.
With a method under development,
the sensitive band should extend to $\sim$4~MeV
(solid blue line in Figure~\ref{fig:ETCC_satellite_performance} {\it left}).
ETCC can constrain SPD below 1\DEGREE for high-energy recoil electrons,
providing sub-degree point source images above $\sim$2~MeV up to 20~MeV 
\citep{Mizumura2018a}.

Gamma-ray photons above 10$-$20~MeV interact with TPC gas 
via electron$-$positron pair production.
ETCC can also detect such events (see Figure~\ref{fig:Readout} {\it left}),
and by improving the triggering algorithm,
the sensitivity should further extend up to $\sim$100~MeV \citep[earlier work was done by][]{Ueno2011a}.
Recoils of nuclei in this process, which cannot be tracked even with ETCC,
are insignificant in gas above $\sim$29~MeV.
The incoming $\gamma$-ray photons between 29$-$100~MeV should be 
constrained better than $\sim$1\DEGREE.

ETCC can also make the best polarization measurements in the MeV band.
Since it constrains the scattering direction of every $\gamma$-ray photon,
it can naturally measure polarizations of any $\gamma$-ray source in its large FOV,
unlike conventional Compton $\gamma$-ray polarimeters, which need to
narrow incoming $\gamma$-ray radiation with collimators \citep{Komura2017a}.
Furthermore, ETCC's powerful background rejection capability will bring
high-quality polarization data in intense background conditions in space,
which should help detect the polarization of faint and/or weakly polarized sources.

\begin{figure}[t]
\vspace*{-0.9in}
\includegraphics[width=6.7in]{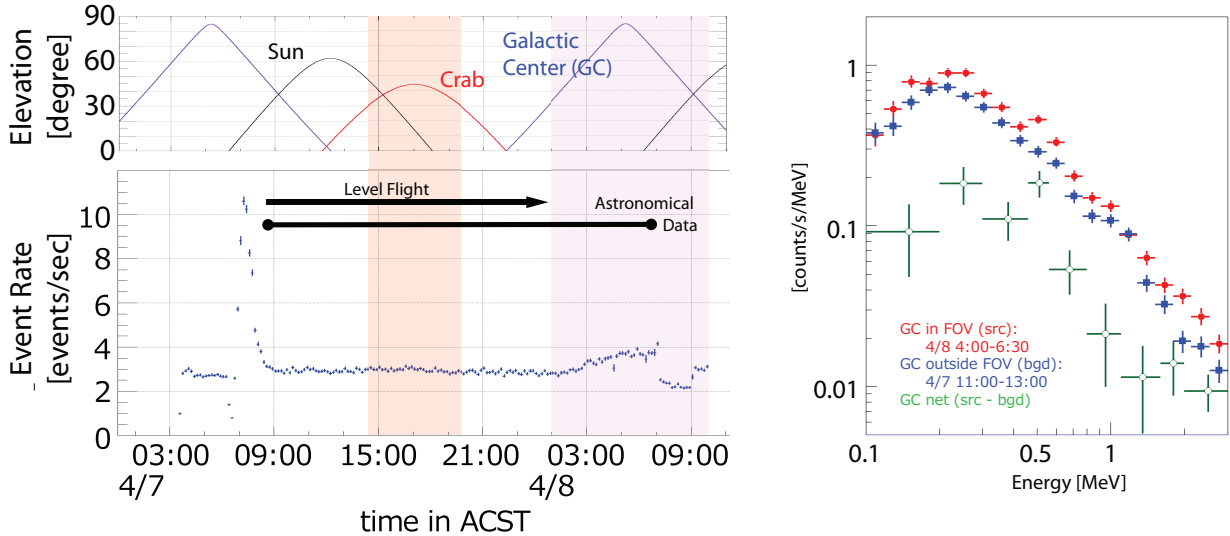}
\caption{Result of the SMILE-2+ balloon flight in 2018 April.
{\it Top left} --- Elevations of the important objects during the flight.
{\it Bottom left} --- Event rate after all event screenings. 
The event rate clearly increased as the Galactic Center region gradually come into the FOV ($<$60\DEGREE from zenith),
indicating that particle background was excluded significantly.
The event rate declined after April 8 8~am when the instrument picked up the noise.
{\it Right} --- Galactic Center spectrum after subtracting a spectrum without the Galactic Center in the FOV as background.
\label{fig:Balloon_Flight}
}
\end{figure}

\subsection{Successful Astronomical Observations in 2018 April}

SMILE-2+ made a 1-day balloon flight in 2018 April from the NASA balloon site near Alice Springs Australia.
The flight reached an altitude of $\sim$39~km, enabling astronomical observations for 22 hours 
(Figure~\ref{fig:Balloon_Flight} {\it bottom left}).
During the flight, TPC always faced the zenith.
Detailed data analyses are still underway, but preliminary results have already demonstrated significant detections of
MeV $\gamma$-ray emission from celestial sources 
at levels that took previous space observatories multiple years to achieve.

Figure~\ref{fig:Readout} {\it right} shows energy histograms of event counts of the flight data after individual screening steps.
The plot displays the effect of each screening.
After the screening, background contamination is reduced by two orders of magnitude \citep[earlier measurement was done by][]{Takada2011a}.
Figure~\ref{fig:Balloon_Flight} {\it bottom left} shows the event rate after all screenings.
A notable variation occurred after midnight; the event rate gradually increased by $\sim$30\%
as the Galactic Center region slowly moved into the ETCC FOV center (i.e., zenith).
This increase can be explained by the flux difference between the Galactic Center region (Galactic Diffuse MeV $\gamma$-rays)
and the extragalactic region (Cosmic Background MeV $\gamma$-rays) seen in earlier measurements \citep{Ajello2008a,Ackermann2015a}
if particle background remained only a few 10\% of the total signal.
This means that ETCC removed particle background events very efficiently
under a high background environment.

We extracted a Galactic Center spectrum by subtracting data without the Galactic Center
from ones with the Galactic Center (Figure~\ref{fig:Balloon_Flight} {\it right}).
The Galactic Center was well inside the FOV only for 2.5 hours between April 8 4~am$-$6:30~am during the flight,
but the net spectrum shows significant emission between 0.1$-$2~MeV, as well as
a clear enhancement at $\sim$511~keV from electron-positron annihilation.
The net spectrum did not change remarkably with background in other time intervals.
The detection significance is $>$10$\sigma$ for the Galactic diffuse $\gamma$-ray emission and 
$\sim$5~$\sigma$ for the 511~keV annihilation line.
COSI and \INTEGRAL/SPI also detected the 511~keV line at 5$\sigma$ in 6.1$\times$10$^{5}$~sec\footnote{https://fskbhe1.puk.ac.za/people/mboett/Texas2017/Kierans.pdf} 
and 58$\sigma$ in 2.1$\times$10$^{7}$~sec, respectively \citep{Siegert2016a}.
SMILE-2+ would have detected the line at 20$-$48~$\sigma$ or $\sim$140$\sigma$
at these exposures.
This result shows that SMILE-2+ has much higher sensitivity than the earlier experiments.

SMILE-2+ observed the Crab nebula only at off-center positions for $\sim$3 hours
when the elevation reached $\sim$45\DEGREE.
Nevertheless, the on-source spectrum clearly shows 
an excess between 0.2$-$0.8~MeV compared to off-source spectra.
Since $\gamma$-rays from low elevation sources need large scattering angles to be detected with a PSA,
the recoil electrons tend to obtain larger energies and thus are prone to escape from the TPC.
The new analysis method (see section~\ref{subsec:design_performance}) should be able to recover
the higher energy spectrum.
The obtained Crab spectrum is consistent with the one estimated from the SMILE-2+ design,
confirming the expected performance during the flight.

The SMILE-2+ balloon experiment in 2018 demonstrated that the ETCC can
make reliable and sensitive measurements of the MeV $\gamma$-ray sky
at balloon altitudes even in a high particle radiation environment.

\begin{table}[t]
\begin{center}
\caption{Roadmap of the ETCC Development}\label{tbl:roadmap}
\begin{tabular}{lccccccc}
\hline\hline
Model&Eff&$\Delta E/E^{\ast}$&PSF&Band&FoV&Sensitivity&Year\\
&(cm$^{-2}$)&(\%)&(degree)&(MeV)&(str)&(mCrab)\\ \hline
SMILE-2+&1&12&10&0.2$-$2$^{\dagger}$&3$^{\ddagger}$&100 [1 day] & 2018\\
SMILE-3&10$-$20&8$-$9&5&0.2$-$10&3$^{\ddagger}$&20 [14-50 day] & $\sim$2022\\
ETCC satellite&200&2&2&0.1$-$100&$>$4&1 [1 year] & $\sim$2030\\ \hline
\end{tabular}
\end{center}
$\ast$: at 662~keV.
$\dagger$: Single PSA hit events only. 
$\ddagger$: Atmospheric $\gamma$-ray background is strong toward the horizontal directions at the balloon altitude.
\end{table}

\subsection{Roadmap of the Further ETCC Development}

SMILE-2+ is still a prototype model developed under limited funding and workforce.
The ETCC performance should improve greatly with relatively minor upgrades.
We are currently developing the new ETCC, SMILE-3, which will achieve an effective area of
10$-$20~cm$^{2}$ at 500~keV and a PSF better than 5\DEGREE (FWHM)  (Table~\ref{tbl:roadmap}).
We are proposing a long duration ($\gtrsim$12 days) balloon flight of SMILE-3 from 
New Zealand in 2022, which is expected to achieve $\sim$5 times better sensitivity
than the 8-year COMPTEL survey.
This flight will verify a future space mission, as well as producing important science results.
Details of the ETCC development for SMILE-3 and the proposed space observatory are described below.

The ETCC's effective area is expected to increase by i) enhancing the Compton scattering efficiency in the TPC,
ii) improving the detection efficiency of scattered $\gamma$-rays, and iii) decreasing the instrumental
dead time.
For i),
the new ETCC models use gas with higher molecular weight, which has more electrons per molecule.
The current best candidate is CF$_{4}$, which has 2.3 times more electrons per molecule than Ar.
The gas is pressurized at 3~atm instead of 2~atm for SMILE-2+, which increases the density by 50\%.
Besides, the future models use a 50~cm cube for TPC, which has 4.6 times more physical volume than a 30~cm cube
used for SMILE-2+.
For ii),
we will use a scintillator with a longer radiation length (R.L.),
such as GAGG or LaBr3 with 5 R.L. instead of GSO used for SMILE-2+, which had 1$-$2 R.L.
This increases the triggering efficiency to a few tens of percent.
To detect low energy $\gamma$-ray photons between 0.07$-$0.2 MeV, 
the satellite ETCC can add CdZnTe detectors between the TPC and the PSA,
which are also useful for collecting any recoil electrons, which escape from the TPC (see also the next paragraph).
For iii),
the whole detector is covered with plastic scintillator for vetoing particle background events.
After all these improvements,
the effective area of one ETCC module should be $\sim$50~cm$^{-2}$ at 1~MeV.

The spectral resolution of incoming $\gamma$-rays depends mostly on the energy resolution of the calorimeter.
The recoil electron energy is a fraction (mostly $\lesssim$20\%) of the incoming $\gamma$-ray energy and
is measured well with the current TPC ($\sim$5\% at 100~keV, $\sim$3.5\% at 200~keV).
The SMILE-3 calorimeter plans to use a combination of LaBr3 scintillator plus multi-pixel photon counters (MPPC)
instead of the GSO scintillator and PMT used for SMILE-2+, which should improve the energy resolution from 12\% to 8$-$9\% at 662 keV \citep{Kurosawa2010a}.
However, MPPCs are probably vulnerable to particle radiation in space, so the satellite ETCC calorimeter will use PMTs.
To improve energy resolution below 1~MeV,
the calorimeter will have thick ($\sim$2~cm) CdZnTe detectors with $\sim$1.5 R.L. at 1 MeV (or CdZnTe detectors are layered if necessary).
In this way, the satellite ETCC will achieve an energy resolution of 2$-$3\% between 0.07$-$1~MeV and 2$-$5\% between 1$-$10~MeV. 

\begin{figure}[t]
\includegraphics[width=6.5in]{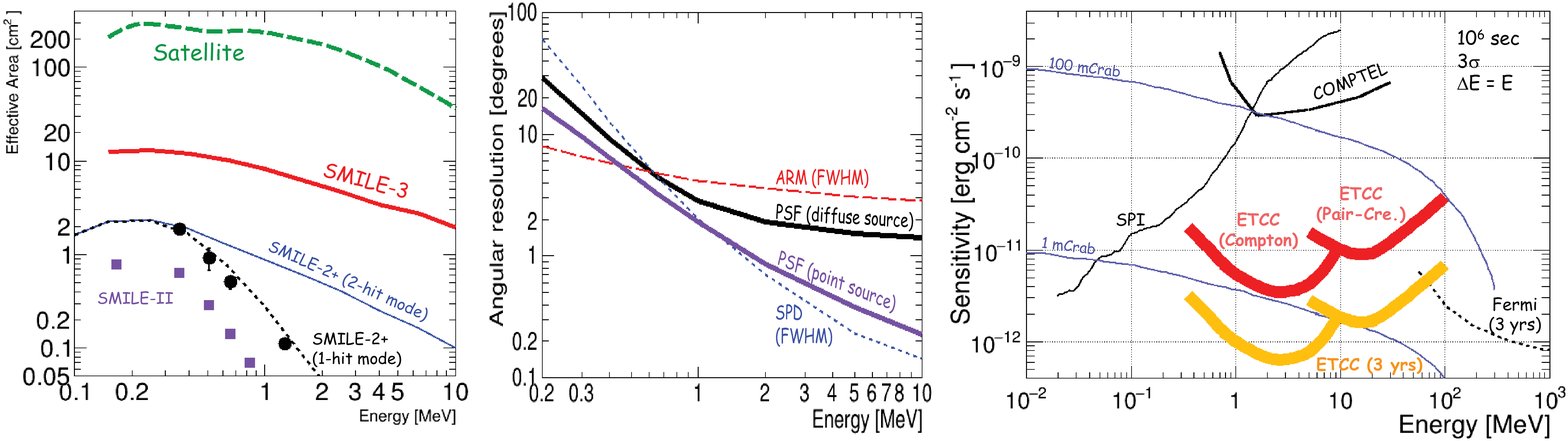}
\caption{
{\it Left} ---  Effective Area of the ETCC models.
{\it Middle} --- Expected angular resolution with CF$_{4}$ gas $+$ LaBr3.
{\it Right} --- Expected sensitivity for 10$^{6}$~sec and 3~years.
\label{fig:ETCC_satellite_performance}
}
\end{figure}

The angular resolution is improved by reducing ARM and SPD.
ARM is directly related to the energy resolutions of the TPC and calorimeter.
With the improvements of the spectral resolution in the previous paragraph,
the angular resolution is expected to be 3$-$4\DEGREE at 662~keV.
As for SPD,
recoil electrons suffer less Coulomb scattering in CF$_{4}$ than Ar.
By using CF$_{4}$, the initial recoil direction should be constrained more precisely.
On the other hand, SMILE-2+ uses orthogonal readout strips for $\mu$-PIC, which introduce ghost images for some tracks \citep{Tanimori2015a}.
The future ETCCs use a 3-axis readout strip system, which substantially suppresses ghost images and 
improves the angular resolution to less than a few tens of degrees.
With these improvements, the angular resolution is expected to improve to sub-degree level 
(see Figure~\ref{fig:ETCC_satellite_performance} {\it middle}).

The timing resolution is limited by the scintillator response, which is about $\mu$sec.
The SMILE-2+ onboard clock only had a millisecond time resolution, but the proposed mission
can decrease time resolution to 1$-$10~$\mu$sec with a minor adjustment of the onboard computer.

In this way, the upgraded ETCCs will satisfy the performance for scientific observing missions (Table~\ref{tbl:roadmap}).
We plan to test the stability of the instrument during a long-duration balloon flight (SMILE-3) expected to be launched in 2022.
SMILE-3 will be equipped with a gas purification system to remove outgassing water
to keep the gas quality during the long-duration flight.
The system was already tested in SMILE-2+ and is expected to work to keep 
the ETCC performance for several months.

\section{A Proposed Space-based MeV Observatory}

We believe that the upgraded ETCCs will revolutionize MeV $\gamma$-ray astronomy especially if they can observe from space for years.
This white paper proposes a space-based mission with ETCCs for an all-sky MeV $\gamma$-ray survey.

ETCC and the \FERMI LAT instrument are $\gamma$-ray observatories with a wide field of views.
ETCC covers $\gtrsim$4 steradian, while \FERMI LAT covers $\sim$20\% of the sky.
We will thus employ an operational strategy similar to the \FERMI observatory.
The ETCC space observatory would fly on a low-earth orbit to minimize particle background radiation.
The ETCC telescope faces away from the Earth at all times, and scans a half of the sky every other satellite orbit,
covering the whole sky every 3 hours.
The spacecraft may have an option of continuous pointings for transient sources,
such as gravitational-wave events and distant gamma-ray bursts.
The spacecraft requires movable solar arrays for observing flexibility.

We aim to achieve an all-sky MeV $\gamma$-ray survey at the sub mCrab level in 3~years,
which is comparable to the \FERMI GeV $\gamma$-ray survey (Figure~\ref{fig:ETCC_satellite_performance} {\it right}).
This goal requires an effective area at $\sim$200~cm$^{-2}$@1~MeV, which can be achieved with $\sim$4 upgraded ETCC modules.
One ETCC module is placed in a tube-shaped chamber with a diameter of $\sim$120~cm and a height of $\sim$80~cm 
(Figure~\ref{fig:ETCC_satellite} {\it left}).
Based on SMILE-2+ ETCC, which uses off-the-shelf commercial supplies,
an ETCC module is estimated to weigh $\sim$350 kg with electrical consumption at 450$-$490~W.
With 4 ETCC modules,
the detector component weighs $\sim$1.4~ton and consumes $\sim$1.9~kW.
These numbers may be reduced if it is optimized for a spacecraft mission.
Based on the \FERMI observatory design,
the satellite bus system is estimated at $\sim$500~kg, so the total weight of the satellite will be $\sim$2~ton.
Placing multiple ETCCs onboard is good for redundancy as well.

Data will be preprocessed onboard to accommodate the available telemetry.
We estimate that ETCC satellite will obtain thousands of counts per second on average based on the SMILE-2+ result.
We expect that an event data can be described with $\sim$100~bytes, so the data rate will be $\sim$9~Gbytes per day.
The onboard computer will monitor transient sources regularly, to quickly alert ground contacts through TDRSS.

Potential launchers are Delta-II or Space-X Falcon 9.
The spacecraft may be folded during a launch to fit inside a fairing.
The developing cost will be comparable to \FERMI, \$500M$-$1B Medium-size mission.

\section{ETCC Advantages over Other MeV Detectors}

ETCCs have demonstrated that tracking Compton recoil electrons is a powerful tool 
for localizing incoming MeV $\gamma$-ray photons and discriminating particle backgrounds.
Electron tracking is arguably the best way to advance the field of MeV $\gamma$-ray astronomy.
Prof. Sch{\"o}nfelder, a leader of the COMPTEL mission, stressed the necessity of instruments with 
electron tracking and ``redundant" measurement of physical values for the next generation 
MeV $\gamma$-ray observatory \citep{Schoenfelder2004a}.

\begin{figure}
\hspace*{0.15in}
\includegraphics[width=6in]{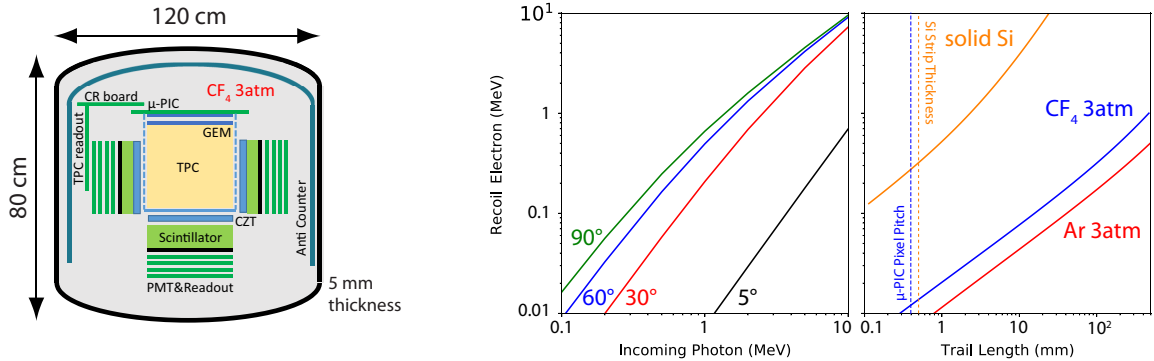}
\caption{
{\it Left} --- Schematic view of an ETCC module for a space observatory.
Multiple ETCC modules may be put into a single pressure vessel to reduce the total weight.
{\it Middle} --- Incoming photon energy vs. Recoiled electron energy for scattering angles, 5\DEGREE, 30\DEGREE, 60\DEGREE
and 90\DEGREE.
{\it Right} --- Trail length of the recoiled electrons vs. Recoiled electron energy for 
CF$_{4}$ gas, Ar gas, and solid Si  (Ref. NIST/ESTAR).
\label{fig:recoiled_electrons_gas_solid}
\label{fig:ETCC_satellite}
}
\end{figure}

Most concept studies of next-generation MeV $\gamma$-ray observatories are based on semiconductor devices.
However, electrons recoiled with Compton scattering do not travel long inside semiconductor materials (Si, Cd etc).
Figure~\ref{fig:recoiled_electrons_gas_solid} {\it middle \& right} show relation between the incident $\gamma$-ray photon energy, 
scattering angle, recoiled electron energy, and electron traveling length in different elements.
For example, 511~keV photons scatter most at $\sim$35\DEGREE according to the Klein-Nishina cross-section.
Their recoil electrons have $E\sim$80~keV and travel $\sim$2.8~cm in Ar gas and 1.1~cm in CF$_{4}$ gas pressurized at 3 atm,
which can be easily tracked with ETCC's $\mu$-PIC with 400~$\mu$m pixel sensors.
On the other hand, these electrons travel only 50~$\mu$m in solid silicon, which cannot be tracked 
with any semiconductor devices currently available.
Semiconductor devices probably need $\gamma$-rays at $\gtrsim$1~MeV to detect a recoil electron at two different positions.

From our experience of the ETCC development,
we believe that sensitive MeV $\gamma$-ray Compton detectors should hold the following mechanisms to reduce particle background.
First, they measure ``redundant" physical values of the scattering process.
Second, the Compton scattering material is made from low $Z$ elements to reduce unwanted 
photo-absorption of soft $\gamma$-ray photons by the scattering material, thereby increasing the Compton scattering efficiency.
Third, the detector design is simple, without need for complicated readout electrode, electronics, 
or cooling bars, which produces additional background events on-orbit, which cannot be removed with veto counters.
COMPTEL actually satisfied these conditions,
or it might not have detected any celestial $\gamma$-ray sources.
Semiconductor MeV $\gamma$-ray Compton telescopes are difficult to satisfy these conditions.

In summary, an ideal MeV $\gamma$-ray Compton telescope should measure all physical parameters 
related to the scattering process with minimum mechanics.
ETCC is unique in that it measures all physical parameters in each Compton scattering process with sub-mm 3D sampling.
Gas detectors are also good in space without suffering major radiation damages, unlike semiconductor detectors.
We may say that ETCCs are the most advanced MeV $\gamma$-ray detectors for decades to come.

\section{An MeV Observatory for the Entire Community}
We believe that a space-based observatory mission with ETCCs dramatically deepens the knowledge of our universe.
If an ETCC satellite is launched in the 2020s,
it will be complementary to \FERMI in GeV $\gamma$-ray, CTA in TeV $\gamma$-ray and \NUS, {\it XRISM} and {\it Athena} in X-rays,
and this combination would cover the high-energy electromagnetic spectrum.
Its all-sky monitoring capability will have good synergies with LIGO, LISA, and LSST.

The $\gamma$-ray community in the US has a long history and experience of space $\gamma$-ray observations with the CGRO and \FERMI observatories.
We think it is best if NASA could lead this spacecraft mission in collaboration with Tanimori et al.\ for the detector development.
The current team is very small although the anticipated mission is large.
We expect all researchers who are interested in the MeV $\gamma$-ray
science will appreciate the clear advantages of an ETCC space observatory mission.


\clearpage

\section*{Endorsement}
Prof. Roland Diehl (MPE): ``very supportive to your experimental proposal and initiative, in general, as a most-promising advance in the key field of nuclear-line and positron annihilation science."

\newcommand\wpwidth{3mm}

\section*{Related Science White Papers}

\noindent 
Marco Ajello et al., ``Supermassive black holes at high redshifts"\vspace{\wpwidth}\\
Marco Ajello et al., ``The MeV Background"\vspace{\wpwidth}\\
Eric Burns et al., ``A Summary of Multimessenger Science with Neutron Star Mergers"\vspace{\wpwidth}\\
Eric Burns et al., ``Gamma Rays and Gravitational Waves"\vspace{\wpwidth}\\
Regina Caputo et al., ``Looking Under a Better Lamppost: MeV-scale Dark Matter Candidates"\vspace{\wpwidth}\\
Stefano Ciprini et al., ``Gravitationally Lensed MeV Gamma-ray Blazars"\vspace{\wpwidth}\\
Mattia Di Mauro et al., ``Prospects for the detection of synchrotron halos around middle-age pulsars"\vspace{\wpwidth}\\
Ryann Foley et al., ``Gravity and Light: Combining Gravitational Wave and Electromagnetic Observations in the 2020s"\vspace{\wpwidth}\\
Chris Fryer et al., ``Core-Collapse Supernovae and Multi-Messenger Astronomy"\vspace{\wpwidth}\\
Dale Gary et al., ``Particle Acceleration and Transport, New Perspectives from Radio, X-ray, and  Gamma-Ray Observations"\vspace{\wpwidth}\\
Joseph Gelfand et al., ``MeV Emission from Pulsar Wind Nebulae: Understanding Extreme Particle Acceleration in Highly Relativistic Outflows"\vspace{\wpwidth}\\
Bruce Grossan et al., ``Measurement of the Optical-IR Spectral Shape of Prompt Gamma-Ray Burst Emission: A Timely Call to Action for Gamma-Ray Burst Science"\vspace{\wpwidth}\\
Sylvain Guiriec et al., ``Gamma-Ray Science in the 2020s"\vspace{\wpwidth}\\
Alice Harding et al., ``Prospects for Pulsar Studies at MeV Energies"\vspace{\wpwidth}\\
Carolyn Kierans et al., ``Positron Annihilation in the Galaxy"\vspace{\wpwidth}\\
Mark L. McConnell et al., ``Prompt Emission Polarimetry of Gamma-Ray Bursts"\vspace{\wpwidth}\\
Elleen Meyer et al., ``Prospects for AGN Studies at Hard X-ray through MeV Energies"\vspace{\wpwidth}\\
Roopesh Ojha et al., ``Neutrinos, Cosmic Rays, and the MeV Band"\vspace{\wpwidth}\\
Elena Orlando et al., ``Cosmic Rays and interstellar medium with Gamma-Ray Observations at MeV Energies"\vspace{\wpwidth}\\
Bindu Rani et al., ``High-Energy Polarimetry - a new window to probe extreme physics in AGN jets"\vspace{\wpwidth}\\
Bindu Rani et al., ``Multi-Physics of AGN Jets in the Multi-Messenger Era"\vspace{\wpwidth}\\
Marcos Santander et al., ``A Unique Messenger to Probe Active Galactic Nuclei: High-Energy Neutrinos"\vspace{\wpwidth}\\
Frank Timmes et al., ``Catching Element Formation In The Act, The Case for a New MeV Gamma-Ray Mission: Radionuclide Astronomy in the 2020s"\vspace{\wpwidth}\\
Tonia Venters et al.,  ``Energetic Particles of Cosmic Accelerators I: Galactic Accelerators"\vspace{\wpwidth}\\
Tonia Venters et al.,  ``Energetic Particles of Cosmic Accelerators II: Active Galactic Nuclei and Gamma-ray Bursts"\vspace{\wpwidth}\\
Zorawar Wadiasingh et al., ``Magnetars as Astrophysical Laboratories of Extreme Quantum Electrodynamics: The Case for a Compton Telescope"\vspace{\wpwidth}\\
Michael Zingale et al., ``MMA SAG: Thermonuclear Supernovae"

\bibliographystyle{aasjournal}
\bibliography{astro}

\begin{thebibliography}{}
\expandafter\ifx\csname natexlab\endcsname\relax\def\natexlab#1{#1}\fi

\bibitem[{{Ackermann} {et~al.}(2015){Ackermann}, {Ajello}, {Albert}, {Atwood},
  {Baldini}, {Ballet}, {Barbiellini}, {Bastieri}, {Bechtol}, \&
  {Bellazzini}}]{Ackermann2015a}
{Ackermann}, M., {Ajello}, M., {Albert}, A., {et~al.} 2015, \apj, 799, 86

\bibitem[{{Ajello} {et~al.}(2008){Ajello}, {Greiner}, {Sato}, {Willis},
  {Kanbach}, {Strong}, {Diehl}, {Hasinger}, {Gehrels}, \&
  {Markwardt}}]{Ajello2008a}
{Ajello}, M., {Greiner}, J., {Sato}, G., {et~al.} 2008, \apj, 689, 666

\bibitem[{{Beacom} \& {Y{\"u}ksel}(2006)}]{Beacom2006a}
{Beacom}, J.~F., \& {Y{\"u}ksel}, H. 2006, \prl, 97, 071102

\bibitem[{{Chiu} {et~al.}(2017){Chiu}, {Boggs}, {Kierans}, {Lowell}, {Sleator},
  {Tomsick}, {Zoglauer}, {Amman}, {Chang}, \& {Chu}}]{Chiu2017a}
{Chiu}, J.~L., {Boggs}, S.~E., {Kierans}, C.~A., {et~al.} 2017, International
  Cosmic Ray Conference, 301, 796

\bibitem[{{Churazov} {et~al.}(2014){Churazov}, {Sunyaev}, {Isern},
  {Kn{\"o}dlseder}, {Jean}, {Lebrun}, {Chugai}, {Grebenev}, {Bravo}, \&
  {Sazonov}}]{Churazov2014a}
{Churazov}, E., {Sunyaev}, R., {Isern}, J., {et~al.} 2014, \nat, 512, 406

\bibitem[{{Diehl} {et~al.}(2014){Diehl}, {Siegert}, {Hillebrandt}, {Grebenev},
  {Greiner}, {Krause}, {Kromer}, {Maeda}, {R{\"o}pke}, \&
  {Taubenberger}}]{Diehl2014a}
{Diehl}, R., {Siegert}, T., {Hillebrandt}, W., {et~al.} 2014, Science, 345,
  1162

\bibitem[{{Horiuchi} \& {Beacom}(2010)}]{Horiuchi2010a}
{Horiuchi}, S., \& {Beacom}, J.~F. 2010, \apj, 723, 329

\bibitem[{{Komura} {et~al.}(2017){Komura}, {Takada}, {Mizumura}, {Miyamoto},
  {Takemura}, {Kishimoto}, {Kubo}, {Kurosawa}, {Matsuoka}, \&
  {Miuchi}}]{Komura2017a}
{Komura}, S., {Takada}, A., {Mizumura}, Y., {et~al.} 2017, \apj, 839, 41

\bibitem[{{Krivonos} {et~al.}(2015){Krivonos}, {Tsygankov}, {Lutovinov},
  {Revnivtsev}, {Churazov}, \& {Sunyaev}}]{Krivonos2015a}
{Krivonos}, R., {Tsygankov}, S., {Lutovinov}, A., {et~al.} 2015, \mnras, 448,
  3766

\bibitem[{{Kurosawa} {et~al.}(2010){Kurosawa}, {Kubo}, {Hattori}, {Ida},
  {Iwaki}, {Kabuki}, {Kubo}, {Kunieda}, {Miuchi}, \&
  {Nakahara}}]{Kurosawa2010a}
{Kurosawa}, S., {Kubo}, H., {Hattori}, K., {et~al.} 2010, Nuclear Instruments
  and Methods in Physics Research A, 623, 249

\bibitem[{{Mizumura} {et~al.}(2018){Mizumura}, {Takada}, \&
  {Tanimori}}]{Mizumura2018a}
{Mizumura}, Y., {Takada}, A., \& {Tanimori}, T. 2018, arXiv e-prints,
  arXiv:1805.07939

\bibitem[{Mizumura {et~al.}(2014)Mizumura, Tanimori, Kubo, Takada, Parker,
  Mizumoto, Sonoda, Tomono, Sawano, Nakamura, Matsuoka, Komura, Nakamura, Oda,
  Miuchi, Kabuki, Kishimoto, Kurosawa, \& Iwaki}]{Mizumura2014a}
Mizumura, Y., Tanimori, T., Kubo, H., {et~al.} 2014, Journal of
  Instrumentation, 9, C05045

\bibitem[{{Nakamura} {et~al.}(2018){Nakamura}, {Tanimori}, {Takada},
  {Mizumura}, {Komura}, {Kishimoto}, {Takemura}, {Yoshikawa}, {Tanigushi}, \&
  {Onozaka}}]{Nakamura2018a}
{Nakamura}, Y., {Tanimori}, T., {Takada}, A., {et~al.} 2018, in Society of
  Photo-Optical Instrumentation Engineers (SPIE) Conference Series, Vol. 10699,
  \procspie, 106995W

\bibitem[{{Sch{\"o}nfelder}(2004)}]{Schoenfelder2004a}
{Sch{\"o}nfelder}, V. 2004, \nar, 48, 193

\bibitem[{{Sch{\"o}nfelder} {et~al.}(1993){Sch{\"o}nfelder}, {Aarts},
  {Bennett}, {de Boer}, {Clear}, {Collmar}, {Connors}, {Deerenberg}, {Diehl},
  \& {von Dordrecht}}]{Schoenfelder1993a}
{Sch{\"o}nfelder}, V., {Aarts}, H., {Bennett}, K., {et~al.} 1993, \apjs, 86,
  657

\bibitem[{{Siegert} {et~al.}(2016){Siegert}, {Diehl}, {Khachatryan}, {Krause},
  {Guglielmetti}, {Greiner}, {Strong}, \& {Zhang}}]{Siegert2016a}
{Siegert}, T., {Diehl}, R., {Khachatryan}, G., {et~al.} 2016, \aap, 586, A84

\bibitem[{{Takada} {et~al.}(2011){Takada}, {Kubo}, {Nishimura}, {Ueno},
  {Hattori}, {Kabuki}, {Kurosawa}, {Miuchi}, {Mizuta}, \&
  {Nagayoshi}}]{Takada2011a}
{Takada}, A., {Kubo}, H., {Nishimura}, H., {et~al.} 2011, \apj, 733, 13

\bibitem[{{Takada} {et~al.}(2016){Takada}, {Tanimori}, {Kubo}, {Mizumoto},
  {Mizumura}, {Komura}, {Kishimoto}, {Takemura}, {Yoshikawa}, \&
  {Nakamasu}}]{Takada2016a}
{Takada}, A., {Tanimori}, T., {Kubo}, H., {et~al.} 2016, in Society of
  Photo-Optical Instrumentation Engineers (SPIE) Conference Series, Vol. 9905,
  \procspie, 99052M

\bibitem[{{Tanimori} {et~al.}(2015){Tanimori}, {Kubo}, {Takada}, {Iwaki},
  {Komura}, {Kurosawa}, {Matsuoka}, {Miuchi}, {Miyamoto}, \&
  {Mizumoto}}]{Tanimori2015a}
{Tanimori}, T., {Kubo}, H., {Takada}, A., {et~al.} 2015, \apj, 810, 28

\bibitem[{{Tanimori} {et~al.}(2017){Tanimori}, {Mizumura}, {Takada},
  {Miyamoto}, {Takemura}, {Kishimoto}, {Komura}, {Kubo}, {Kurosawa}, \&
  {Matsuoka}}]{Tanimori2017a}
{Tanimori}, T., {Mizumura}, Y., {Takada}, A., {et~al.} 2017, Scientific
  Reports, 7, 41511

\bibitem[{{Tomono} {et~al.}(2017){Tomono}, {Mizumoto}, {Takada}, {Komura},
  {Matsuoka}, {Mizumura}, {Oda}, \& {Tanimori}}]{Tomono2017a}
{Tomono}, D., {Mizumoto}, T., {Takada}, A., {et~al.} 2017, Scientific Reports,
  7, 41972

\bibitem[{Tsurutani {et~al.}(2018)Tsurutani, Park, Falkowski, Lakhina, Pickett,
  Bortnik, Hospodarsky, Santolik, Parrot, Henri, \& Hajra}]{Tsurutani2018a}
Tsurutani, B.~T., Park, S.~A., Falkowski, B.~J., {et~al.} 2018, Journal of
  Geophysical Research: Space Physics, 123, 10,009

\bibitem[{{Ueno} {et~al.}(2011){Ueno}, {Tanimori}, {Kubo}, {Miuchi}, {Kabuki},
  {Iwaki}, {Higashi}, {Parker}, {Kurosawa}, \& {Takahashi}}]{Ueno2011a}
{Ueno}, K., {Tanimori}, T., {Kubo}, H., {et~al.} 2011, Nuclear Instruments and
  Methods in Physics Research A, 628, 158

\end{thebibliography}

\end{document}